\documentclass[journal]{IEEEtran}

\usepackage{main_header}

\begin{document}

\title{Quantifying Interference-Assisted Signal Strength Surveillance of Sound Vibrations}

\author{
Alemayehu Solomon Abrar, 
Neal Patwari and 
Sneha Kumar Kasera

\thanks{ Alemayehu Solomon Abrar is with the Preston M. Green Department of Electrical \& Systems Engineering at Washington University in St. Louis}

\thanks{ Neal Patwari is with McKelvey School of Engineering at Washington University in St. Louis}

\thanks{ Sneha Kumar Kasera is with the School of Computing at University of Utah}

}

\maketitle

\begin{abstract} 
A malicious attacker could, by taking control of internet-of-things devices, use them to capture received signal strength (RSS) measurements and perform surveillance on a person's vital signs, activities, and sound in their environment. This article considers an attacker who looks for subtle changes in the RSS in order to eavesdrop sound vibrations.  The challenge to the adversary is that sound vibrations cause very low amplitude changes in RSS, and RSS is typically quantized with a significantly larger step size.  This article contributes a lower bound on an attacker's monitoring performance as a function of the RSS step size and sampling frequency so that a designer can understand their relationship.  Our bound considers the little-known and counter-intuitive fact that an adversary can improve their sinusoidal parameter estimates by making some devices transmit to add interference power into the RSS measurements.  We demonstrate this capability experimentally.  As we show, for typical transceivers, the RSS surveillance attacker can monitor sound vibrations with remarkable accuracy. New mitigation strategies will be required to prevent RSS surveillance attacks.

\end{abstract}

\begin{IEEEkeywords}
Received signal strength, respiratory rate monitoring, sound eavesdropping
\end{IEEEkeywords}

\section{Introduction}
\label{sec:intro}

\IEEEPARstart{E}{xisting} internet-of-things (IoT) devices are notoriously easy to compromise \cite{ifsec, Antonakakis2017understanding, reaper}. Given that devices bring sensors like microphones and cameras into our private spaces \cite{warren2017amazon}, people are rightfully concerned for their privacy.  People know what kind of information an attacker could obtain from compromising a video camera in their private spaces, and may not deploy them \cite{townsend2011privacy} purely due to privacy concerns.  Some may consider themselves at high risk for attacks to their privacy and may physically disable a video camera, like Facebook CEO Mark Zuckerberg \cite{warren2017amazon}. Even among the privacy conscious, though, there is little awareness of what an attacker could obtain from compromising a device which can measure received signal strength (RSS).  Yet, \emph{every} wireless device could be a radio frequency (RF) sensor. 

\begin{figure}[tbhp]
\centering
     \includegraphics[width=0.9\columnwidth]{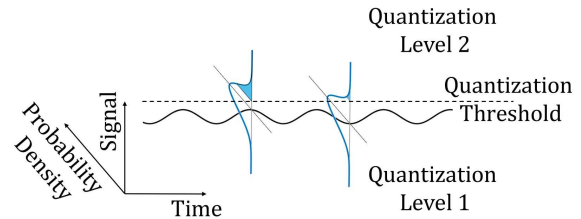}
\caption{Additive noise helps small amplitude sine wave cross a quantization threshold}
\label{fig:intro}
\end{figure}

RF sensors have been shown to be capable of monitoring breathing and heart rate \cite{liu2015tracking}, location \cite{wang2017csi,patwari2010rf}, activity \cite{wang2015understanding}, gesture \cite{pu2013whole}, audio \cite{wei2015acoustic}, and keystrokes \cite{ali2015keystroke}.  An attacker who could remotely control IoT devices would be able to surveil and record data on anyone who is near those devices. One reason why RF sensing surveillance attacks are not addressed in the literature is that RF sensing attackers are seen as having two less-than-ideal choices.  First, an attacker could exploit a device's channel state information (CSI) which can only be obtained from a select group of WiFi network interface cards (NICs). The most capable RF sensing systems reported in the literature use CSI measurements \cite{liu2015tracking,wang2017csi,wang2015understanding, ali2015keystroke}. IoT device makers can avoid this class of attacks simply by avoiding using these NICs in their products.  Second, an attacker can exploit a transceiver's RSS data.  RSS is available almost universally in wireless transceivers because network functions such as multiple access control and power control \cite{baccour2012radio, kim2006cara,lin2006atpc} require it.
RF sensors using RSS have been used to perform contact-free vital sign monitoring \cite{ravichandran2015wibreathe}, device-free localization \cite{youssef2007challenges}, and gesture and activity recognition \cite{sigg2014telepathic} in limited settings. RSS is thought to be coarse and limiting because low-amplitude changes in RSS due to sound or pulse vibrations for example, can easily be lost due to the large quantization step size of most RSS measurements.
This article shows for the first time that this intuition is not true.  As we show, an attacker can use an extra compromised transceiver to transmit what we call \emph{helpful interference} (HI).  We describe the counter-intuitive idea that some extra noise in the received power, due to an interferer's signal, will enable reliable estimation of the low-amplitude changes to the received power, even with large quantization step sizes in the RSS. In addition to demonstrating HI, this article addresses the question, what is the best that an attacker could possibly do? We present an analytical bound on the best performance for the attacker in this scenario.

Our bound has two purposes.  Knowing an attacker's limits can provide a guarantee to a user, that even if the device is completely compromised by an attacker, its ability has a particular quantitative limit.  Furthermore, if these limits are known as a function of transceiver parameters, then a transceiver designer can adapt the design to reduce an attacker's capabilities.  
\begin{figure*}[tbhp]
\begin{center}
   \begin{subfigure}[t]{0.45\textwidth}
     \includegraphics[width=\textwidth]{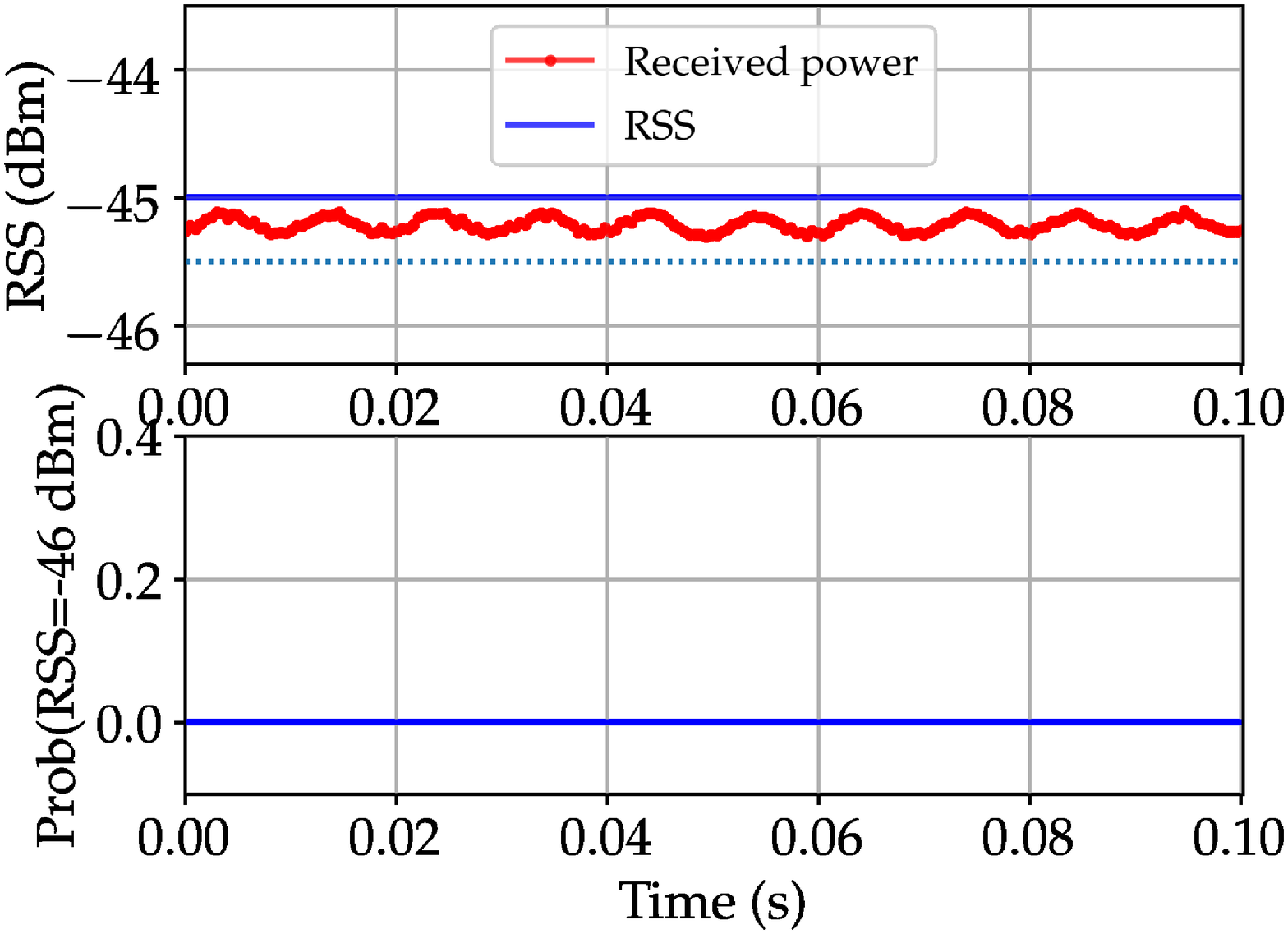}
     \caption{}
     \label{fig:intoff}
  \end{subfigure} %
  \qquad
  %
  \begin{subfigure}[t]{0.45\textwidth}
     \includegraphics[width=\textwidth]{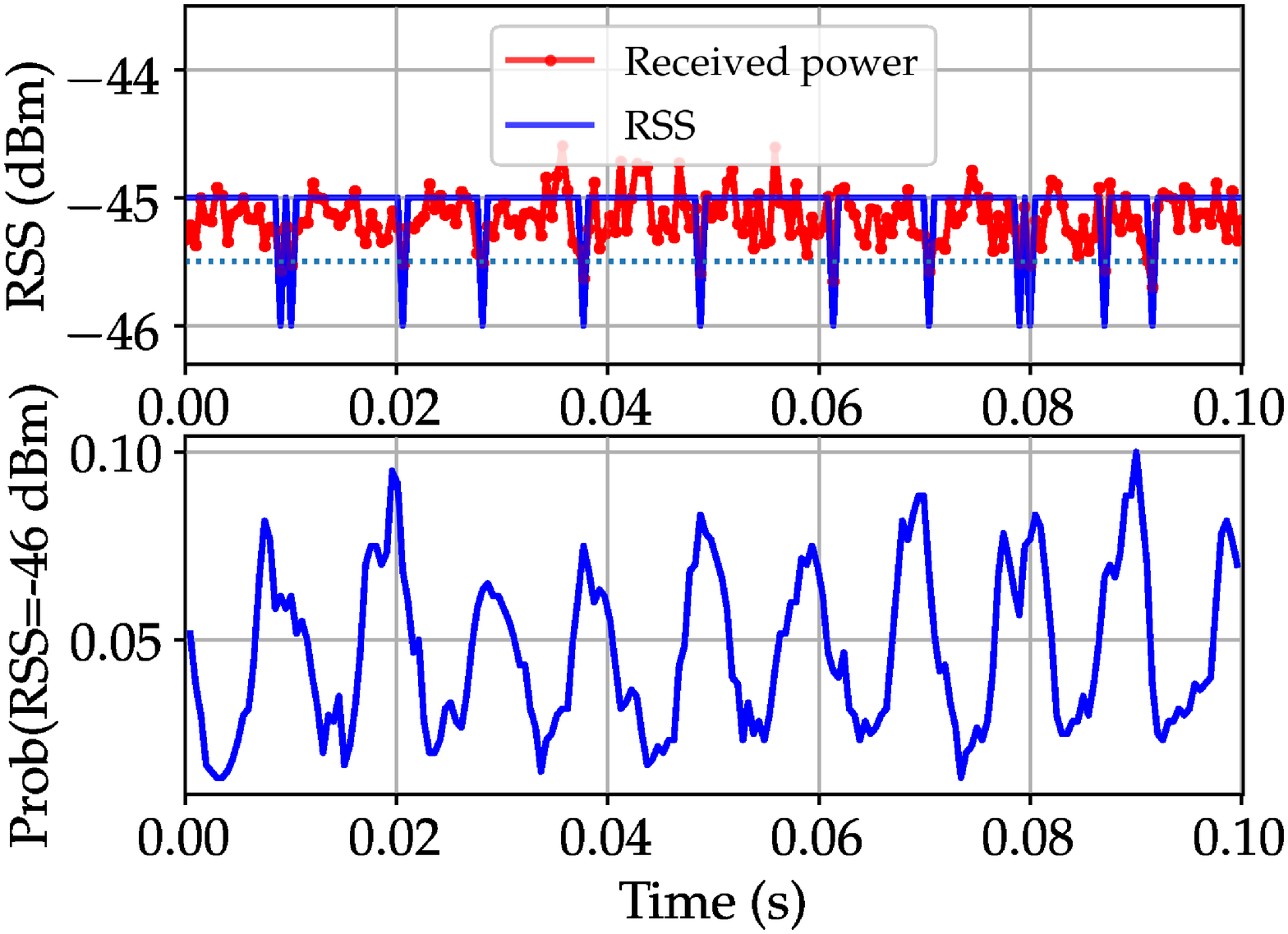}
     \caption{}
     \label{fig:quantization}
  \end{subfigure}
\end{center}
  \caption{Received power with RSS quantized to the nearest integer and spectrogram of the RSS, while a 100 Hz single tone sound is played on Google Home speaker.  (a) Without interference, RSS is constant, but (b) with interference the RSS crosses to the lower value at least once per period, enabling spectral estimation of the sound. }
 \label{fig:noise_rss}
 \end{figure*}
 
\vspace{0.1in} \noindent {\bf Quantization and Interference}: Quantization is generally thought to be good news for security against a privacy attacker with access to RSS. A well-known limitation of RSS is that it is quantized, typically with 1 dB step size (although sometimes 0.5 dB or 4 dB). Typical changes to received power due to sound are much less than 1 dB. Fig.~\ref{fig:intoff}  shows what often happens---the received power (displayed as $\textcolor{red}{---}$) is affected by vibrations caused by a 100 Hz single-tone sound played on a speaker, but the quantized RSS (displayed as $\textcolor{blue}{---}$) is constant.  
However, the bad news we discover in this investigation is that an attacker's capabilities are greater than previously thought because the attacker can exploit what we call \emph{helpful interference} (HI).  HI is the purposeful transmission of interference from other devices to increase the variance of the RSS measurement at a receiver.  An attacker could use other compromised devices to transmit, perhaps with carrier sense disabled, to generate HI.  Counterintuitively, this increase in measurement variance prior to quantization can actually improve the attacker's estimates of frequency and amplitude, partially negating the effects of quantization.  Fig.~\ref{fig:quantization} shows an example of how noise from an interferer helps to sometimes ``push'' the quantized RSS over the boundary to the next RSS value, thus making the 100 Hz sound observable on the RSS spectrogram. We provide the first experimental demonstration, to our knowledge, of the ability to use received power from commercial-off-the-shelf (COTS) transceivers to record sound and the use interference to improve the performance of an RSS-based sound eavesdropping. Our experimental observations with varying levels of helpful interference also exhibit an optimal level of HI.  The estimation bounds presented in this paper take into account an attacker's ability to use HI, and also show that there is an optimal level of HI beyond which performance degrades.  The resulting variance bounds are a function of the transceiver's RSS quantization step size and sampling rate.  Device makers can use this bound to limit the inadvertent measurement capabilities of attackers by the design of their device.

\vspace{0.1in} \noindent {\bf Sound Eavesdropping}:  
We pay particular attention to RSS-based sound eavesdropping in this paper because sound causes only slight changes in RSS while possessing vital private information about a person's activity and their surroundings.  We believe it is important to know the relationship between the performance of sound eavesdropping and quantization step size and RSS sampling rate, which can be computed from the bound in this article.  Further, sound reveals private information about person’s activity, their communications, and objects and incidents in their surroundings. Typically, a person would not want anyone outside of the room to have such data. While we emphasize on sound eavesdropping in this article, the same principle applies to many other RF sensing applications including breathing and heart rate monitoring.

\vspace{0.1in} \noindent {\bf Contribution Summary}: We summarize the contributions of this article as follows:
\begin{enumerate}
    \item We present the first RSS-based sound eavesdropping demonstration using COTS transceivers.
     \item This article is the first to propose, quantify, and experimentally validate the use of interference to \emph{improve} RSS-based sound eavesdropping.
     
    \item This article also provides a lower bound on estimation variance for frequency and amplitude estimates, and applies it to provide quantitative lower limits for RSS-based sound eavesdropping.

\end{enumerate}

In combination, this article shows that an RF sensing surveillance is a greater threat than previously thought.  In particular, while RSS was thought of as a poor choice for an attacker, that helpful interference can increase the information available to the attacker using quantized RSS.  Since RSS is readily available from almost all wireless interfaces, even IoT devices without sensors (e.g., smart light bulbs) can be used for such an attack.  
\section{Threat Model} 
\label{sec:threat} 
We assume that a home has two or more transceivers, and that there are sound vibrations in the vicinity of these devices. 
Slight motions caused by sound vibrations from a speaker or nearby object cause changes in the radio channel that can be observed at the receivers as variations in the received power, and indirectly in RSS, which is the quantized received power. 

We assume that an attacker can access the transmitters and receivers in the home, and that they can alter device firmware or software to force transmitters to transmit more often and  receivers to receive (and collect RSS measurements) more often, up to the maximum rate and maximum RSS precision as possible with the receiver hardware. 
Since wireless standards (e.g, 802.11) require higher layer access to signal strength \cite{sjoberg2010measuring}, an attacker can use this information maliciously for RSS surveillance.
This attack model is practical as it has been shown that there are millions of vulnerable and unprotected Internet connected devices deployed today, and attackers have repeatedly managed to remotely take over such devices and install botnets on them \cite{Antonakakis2017understanding, reaper} or make modifications to the software/firmware \cite{bazhaniuk2018remotely}. 
Furthermore, we assume that an attacker can force a transmitter to transmit in the same frequency band at the same time as the other transmitter (e.g., by disabling carrier sensing \cite{xu2005feasibility, vanhoef2014advanced}, using a hidden terminal) in order to contribute noise to the receiver, as described in \S \ref{sec:eval}.

The attacker can either transfer the measurements to another processor or process the measurements locally on the same device.  We do not assume any computational or communication constraints for the attacker. We make no assumption about the algorithm used except that it is unbiased, e.g., the attacker does not always guess the same breathing rate regardless of the data.

We do not consider an adversary that brings their own wireless devices to the home. While an attacker who brings a software-defined radio (SDR) to a home might be able to monitor a resident with greater accuracy, this would require physical proximity to each home to be attacked and considerable cost for each SDR.  In contrast, the attack we study requires only remote access to the already installed but compromised commercial wireless devices, and thus could be launched without new hardware and on a very large scale.

\section{Effect of Sound on RSS}
\label{sec:sound}
It is counterintuitive that sound would have an effect on measurements of signal strength.  The amplitude of vibrations due to sound (on the order of a millimeter or less) are a small fraction of the wavelength of the RF wave (on the order of 100 mm or higher).  In this section, we explain how a receiver measuring signal strength is capable of measuring sound vibrations that displace a surface in a sinusoidal pattern with a peak-to-peak displacement amplitude of $\Delta z$. 

First, we note that radio waves which reflect or scatter from a vibrating surface will arrive at the receiver with varying phase due to the change in path length due to the movement of the surface. The amplitude of the wave is largely unchanged due to vibration, since the length change is very small compared to the total path length.  However, in a multipath channel, many waves will not be changed by the vibrating surface.  

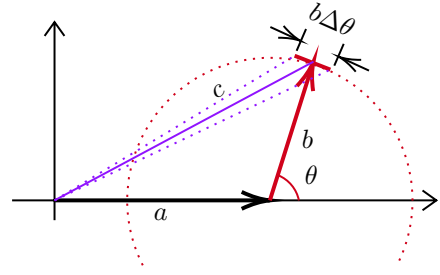
\begin{figure}
    \centering
    \tikzset{every picture/.style={line width=0.75pt}} 

\begin{tikzpicture}[x=0.75pt,y=0.75pt,yscale=-1,xscale=1]

\draw  (39,152) -- (253,152)(60.4,62) -- (60.4,162) (246,147) -- (253,152) -- (246,157) (55.4,69) -- (60.4,62) -- (65.4,69)  ;
\draw [line width=1.5]    (60.4,152) -- (166,152) ;
\draw [shift={(169,152)}, rotate = 180] [color={rgb, 255:red, 0; green, 0; blue, 0 }  ][line width=1.5]    (14.21,-4.28) .. controls (9.04,-1.82) and (4.3,-0.39) .. (0,0) .. controls (4.3,0.39) and (9.04,1.82) .. (14.21,4.28)   ;

\draw [color={rgb, 255:red, 208; green, 2; blue, 27 }  ,draw opacity=1 ][line width=1.5]    (169,152) -- (190.1,84.86) ;
\draw [shift={(191,82)}, rotate = 467.45] [color={rgb, 255:red, 208; green, 2; blue, 27 }  ,draw opacity=1 ][line width=1.5]    (14.21,-4.28) .. controls (9.04,-1.82) and (4.3,-0.39) .. (0,0) .. controls (4.3,0.39) and (9.04,1.82) .. (14.21,4.28)   ;

\draw [color={rgb, 255:red, 208; green, 2; blue, 27 }  ,draw opacity=1 ][line width=1.5]    (181.5,79) -- (199.5,86) ;

\draw  [draw opacity=0][dash pattern={on 0.84pt off 2.51pt}] (104.92,184.76) .. controls (99.81,174.78) and (96.96,163.46) .. (97.05,151.47) .. controls (97.34,111.74) and (129.79,79.76) .. (169.53,80.05) .. controls (209.26,80.34) and (241.24,112.79) .. (240.95,152.53) .. controls (240.87,163.48) and (238.35,173.85) .. (233.9,183.11) -- (169,152) -- cycle ; \draw  [color={rgb, 255:red, 208; green, 2; blue, 27 }  ,draw opacity=1 ][dash pattern={on 0.84pt off 2.51pt}] (104.92,184.76) .. controls (99.81,174.78) and (96.96,163.46) .. (97.05,151.47) .. controls (97.34,111.74) and (129.79,79.76) .. (169.53,80.05) .. controls (209.26,80.34) and (241.24,112.79) .. (240.95,152.53) .. controls (240.87,163.48) and (238.35,173.85) .. (233.9,183.11) ;
\draw  [draw opacity=0] (173.27,139.36) .. controls (178.54,141.57) and (182.49,146.28) .. (183.65,152.01) -- (166.5,155.5) -- cycle ; \draw  [color={rgb, 255:red, 208; green, 2; blue, 27 }  ,draw opacity=1 ] (173.27,139.36) .. controls (178.54,141.57) and (182.49,146.28) .. (183.65,152.01) ;
\draw [color={rgb, 255:red, 144; green, 19; blue, 254 }  ,draw opacity=1 ]   (60.4,152) -- (191,82) ;

\draw [color={rgb, 255:red, 144; green, 19; blue, 254 }  ,draw opacity=1 ] [dash pattern={on 0.84pt off 2.51pt}]  (60.4,152) -- (181.5,79) ;

\draw [color={rgb, 255:red, 144; green, 19; blue, 254 }  ,draw opacity=1 ] [dash pattern={on 0.84pt off 2.51pt}]  (60.4,152) -- (199.5,86) ;

\draw    (169.25,64.5) -- (182.68,70.67) ;
\draw [shift={(184.5,71.5)}, rotate = 204.66] [color={rgb, 255:red, 0; green, 0; blue, 0 }  ][line width=0.75]    (10.93,-3.29) .. controls (6.95,-1.4) and (3.31,-0.3) .. (0,0) .. controls (3.31,0.3) and (6.95,1.4) .. (10.93,3.29)   ;

\draw    (182.5,76.5) -- (187.25,65.5) ;

\draw    (205.57,80.33) -- (219,86.5) ;

\draw [shift={(203.75,79.5)}, rotate = 24.66] [color={rgb, 255:red, 0; green, 0; blue, 0 }  ][line width=0.75]    (10.93,-3.29) .. controls (6.95,-1.4) and (3.31,-0.3) .. (0,0) .. controls (3.31,0.3) and (6.95,1.4) .. (10.93,3.29)   ;
\draw    (201.5,84) -- (206.25,73) ;

\draw (114,159) node   {$a$};
\draw (188,121) node   {$b$};
\draw (200.5,60.5) node [rotate=-23.94]  {$b\Delta \theta $};
\draw (143,97) node [rotate=-329.28]  {$c$};
\draw (190,139) node   {$\theta $};

\end{tikzpicture}
    \caption{Contribution from unaffected waves $a$ and affected waves $b$ add in a phasor sum. The phase $\theta$ changes, tracing an arc length $b\Delta \theta$, and changing the amplitude of the combined RF signal $c$.}
    \label{F:Sum}
\end{figure}

The multipath waves' effects add together as a phasor sum at the receive antenna.  Grouping the phasor sum of waves \emph{not} changed by the vibrating surface as $a$, and grouping the wave(s) affected by the vibration as $b$, we graphically show the phasor sum in Fig. \ref{F:Sum}.  As $\theta$, the phase of $b$, changes with a peak-to-peak phase change of $\Delta\theta$, the $b$ phasor traces an arc length $b\theta$, and the amplitude of the sum $c$ changes. 
The peak-to-peak phase change is at most $4\pi \Delta z / \lambda$ where $\lambda$ is the wavelength of the RF signal. A wave reflecting off of the vibrating surface must travel to the surface and back, doubling the vibration displacement if it travels perpendicularly with respect to the surface.  

The primary question is, is the change in amplitude of the RF signal measurable in the RSS?  The baseline RSS in dB is, $ P = 10 \log_{10} |c|^2$, and using the law of cosines to formuate $c$,
\begin{equation}\label{E:dBPower}
P = 10 \log_{10} \left( 
       a^2 + b^2 + 2ab\cos \theta
    \right).
\end{equation}

\begin{figure}
    \centering
    \includegraphics[width=0.8\columnwidth]{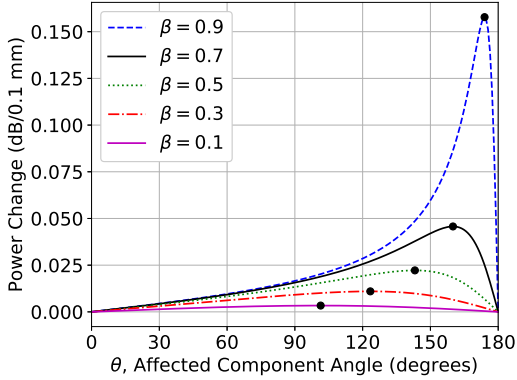}
    \caption{The power change $\Delta P$ vs.\ relative angle of the affected component, for $\Delta z = 0.1$mm, with $\bullet$ showing maximum at $(\theta_{max}, \Delta P_{max})$.}
    \label{F:DeltaP_vs_theta}
\end{figure}

Since the displacement is so small for sound vibrations compared to the wavelength, the change in $\theta$ is small, and we can use a first-order Taylor series approximation to describe the change in power $\Delta P$ as a function of the change in $\theta$.
\begin{equation}
    \Delta P \approx 
    \Delta \theta \left| \frac{\partial P}{\partial \theta} \right|,
\end{equation}
where $\Delta \theta = 4\pi \Delta z / \lambda$ and $P$ is from (\ref{E:dBPower}).  Defining the relative amplitude of the affected component as $\beta = b/a$, the power change becomes,
\begin{equation} \label{E:DeltaP_vs_theta}
    \Delta P \approx 
    \frac{80 \pi \Delta z}{(\ln 10)\lambda}
    \left(\frac{\beta \sin \theta}{1+\beta^2 +2\beta\cos\theta }\right).
\end{equation}
This change in power is plotted in Fig. \ref{F:DeltaP_vs_theta} for a few values of $\beta$ over the range $0 < \theta < \pi/2$ rad.  Here we use a frequency of 915 MHz and $\Delta z = 0.1$mm, but the result is proportional to $\Delta z$ so we explain it as the power change in dB \emph{per} 0.1 mm of peak-to-peak displacement.

We can see there is a maximum possible $\Delta P$ for any value of $\beta$.  Taking the derivative of (\ref{E:DeltaP_vs_theta}) with respect to $\theta$ and setting it to zero, we find the optimal angle of the affected component to be,
\begin{equation}
    \theta_{max} = \arg \max_\theta \Delta P = \cos^{-1}\left(  \frac{-2\beta}{1+\beta^2}\right).
\end{equation}
Next, we assume that $0 < \beta < 1$ because $\beta = b/a$ and we would assume that the fact that $b$ includes a reflection or scattering from the vibrating surface would mean that it would have a lower amplitude than a line-of-sight component.  Using this angle in (\ref{E:DeltaP_vs_theta}), we find that the maximum $\Delta P$ is given by
\begin{equation}
    \Delta P_{max} = \max_\theta \Delta P = 
    \frac{80 \pi \Delta z}{(\ln 10)\lambda}
    \frac{\beta}{1-\beta^2},
\end{equation}
for $0 < \beta < 1$.  These maxima are plotted in Fig. \ref{F:DeltaP_vs_theta}.

Further, if the relative phase $\theta$ is uniformly random between 0 and $2\pi$, we can calculate the expected value of $\Delta P$.  Integrating the product of $\frac{1}{2\pi}$ and $\Delta P$ from (\ref{E:DeltaP_vs_theta}) for $\theta$ between 0 and $2\pi$, we find:
\begin{equation} \label{E:ExpectedDeltaP}
\mbox{E}_\theta[\Delta P] = \left(\frac{8\Delta z}{\lambda}\right) 10\log_{10} \frac{1+\beta}{1-\beta}.
\end{equation}

\paragraph{Discussion}
For very low $\beta$, that is, when the amplitude of the affected component is relatively small, the change in power is approximately proportional to $\beta$ and thus small as well.  It is maximized, however, at an angle of $\theta = \frac{\pi}{2}$ radians or 90$^o$.  

For $\beta$ close to 1, that is, when the affected component has almost the same amplitude as the unaffected component, the change in power can be very high, in fact, $\Delta P$ asymptotically approaches $\infty$ as $\beta \rightarrow 1$.  This behavior means that, in situations when $\beta$ is close to 1 and $\theta$ happens to be a little less than $\pi$, the $\Delta P$ can be very high.  

\begin{figure}
    \centering
    \includegraphics[width=0.8\columnwidth]{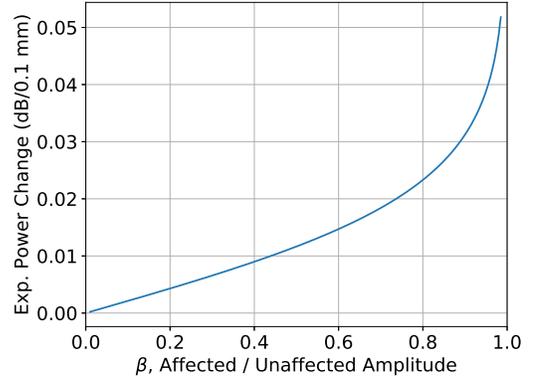}
    \caption{For each 0.1 mm of peak-to-peak displacement due to vibration, the expected value of the change in received power, vs.\ $\beta = b/a$, the relative power in the affected component.}
    \label{F:ExpectedDeltaP}
\end{figure}

Finally, (\ref{E:ExpectedDeltaP}) gives a straightforward formula for finding the expected value of $\Delta P$.  This is plotted in Fig. \ref{F:ExpectedDeltaP} for $\Delta z = 0.1$ mm.  That is, for every 0.1 mm of peak-to-peak displacement due to vibration, we can expect the received power to change as given in Fig. \ref{F:ExpectedDeltaP}.

The results show that measurable power changes should be expected from vibrating objects as long as the amplitude of the affected component is within a few dB of the unaffected component.  At a relative affected power of -6 dB ($\beta = 0.5$), $\mbox{E}[\Delta P] = 0.0116$ dB. This is approximately the same as the standard deviation of error in a single RSS measurement \cite{luong2016rss}.  However, at -20 dB ($\beta = 0.1$), $\mbox{E}[\Delta P] = 0.0021$ dB, about 6 times smaller than the standard deviation of the measurement.  Thus we can see how it is important in these monitoring applications to design the system to keep the amplitude of the affected component within the same order of magnitude as the unaffected component.  

The behavior of power change vs.\ frequency may be more complicated than revealed by (\ref{E:DeltaP_vs_theta}) - (\ref{E:ExpectedDeltaP}).  As given, the power change due to vibration is proportional to the center frequency of the RF signal.  We test a system at 915 MHz, has lower power change by a factor of $2.4/0.915$ compared to a 2.4 GHz system.  However, loss through walls increases dramatically with frequency, for example, showing a linear increase in attenuation with frequency \cite{zhang1994measurements}.  This loss could reduce $b$ and thus $\beta$ in through-wall systems.

While the change in received power due to sound vibrations is mostly in the order of 0.01dB, most wireless transceivers provide received signal strength typically quantized to 1 dB.  In subsequent sections, we present a novel approach to overcome RSS quantization for RSS-based sound eavesdropping.
\section{RSS Surveillance with Helpful Interference}
\label{sec:eval}

In this section, we describe and demonstrate RSS-based sound eavesdropping with helpful interference. To the best of our knowledge, we are the first to introduce the use of helpful interference, that is, transmitting interference to overcome the limitation of RSS quantization in spectral estimation.

\subsection{Devices and Setup}

For our evaluation, we desire a commercial wireless transceiver, but we need control over the quantization step size and sampling rate. We achieve this goal by using a commercial wireless transceiver, the TI CC1200. The CC1200 radio is used as integral part of some commercial internet-of-things (IoT) products \cite{zoul} and it has been shown to measure received power with error only within 0.01dB  \cite{luong2016rss}.
We then can apply any quantizer to the received power in post-processing to emulate the RSS that would have been reported using an arbitrary transceiver RSS quantization step size.

We use CC1200 transceivers configured as transmitter, receiver and HI transmitter nodes. In this experiment, the CC1200 registers are configured with 802.15.4g radio settings at a less congested ISM band at 915 MHz to have a better control on the level of interference in the channel.  While we believe that uncontrolled interference could also benefit sound eavesdropping, we control our interference source for purposes of understanding the relationship between interference power and monitoring performance.  For simplicity, the transmitter sends a continuous wave signal at a transmit power of 12 dBm. The receiver node uses the average of the squared magnitude to compute the received power. This outputs a received power measurement at a rate of about 2 kHz. 
The third transceiver is programmed to generate HI in which we implement a 2-FSK transmitter with a symbol rate of 256 Kbps in the same band and transmit random symbols. To study the effect of the magnitude of interference, we also control the output power of the interferer by changing the value of the \texttt{PA\_CFG1} register on the CC1200 transceiver. 

\begin{figure}[t]
\begin{minipage}[b]{1.0\linewidth}
\centerline{\includegraphics[width=0.95\columnwidth]{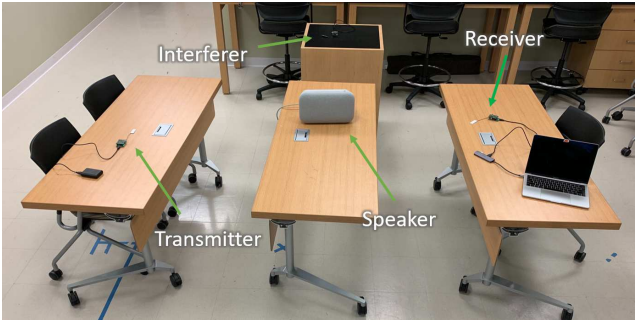}}
\end{minipage}
\caption{Experiment Setup}
\label{fig:setup}
\end{figure}

The experiments were conducted mainly in laboratory settings in a building at Washington University in St. Louis. We run the experiments with three CC1200 transceivers operating as a transmitter, a receiver, and a helpful interference transmitter (HI TX). The transmitter and receiver are typically separated by 2 meters from each other. A Google Home Max speaker is set on a separate table between the transceivers to play audio. Fig. \ref{fig:setup} shows a sample setup used in in this experiment for sound monitoring. 
To evaluate the performance of estimation of the frequency of single-tone sound, we use root mean squared error (RMSE) as an error metric

\subsection{Helpful Interference}
\begin{figure}[t]
\begin{minipage}[b]{1.0\linewidth}
\centerline{\includegraphics[width=0.95\columnwidth]{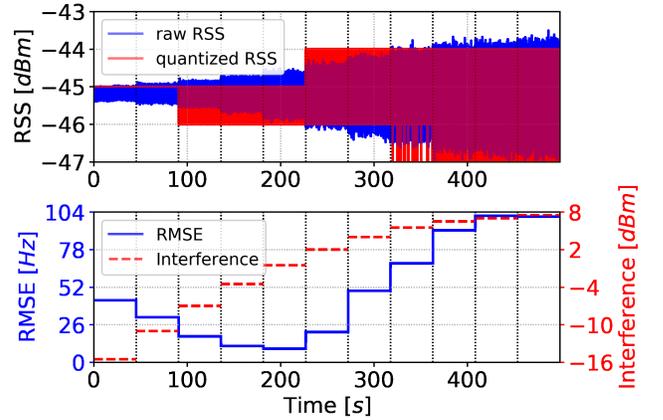}}
\end{minipage}
\caption{As HI power increases each 45s, the frequency estimation error decreases to a minimum of 9 Hz.}
\label{fig:interference}
\end{figure}

We evaluate the effect of increasing interference power on frequency estimation of single-tone sound from quantized RSS measurements. In Fig. \ref{fig:interference}, we show how RSS changes in the presence of increasing interference.  With -15 dBm of interference at the start of the experiment, the measured RSS almost always takes the value of -45 dBm.  As a result, it is largely unable to estimate sound frequency, and the RMSE of the frequency estimate is about 44 Hz.  Each 45 seconds, the HI power is increased by changing the value of the \texttt{PA\_CFG1} register on the interferer radio, as shown in red in the bottom of Fig. \ref{fig:interference}.  As the interferer's power increases, the samples of quantized RSS begin to take more than one different RSS values, initially taking two values at -45 and -46 dBm. This then allows estimating the periodicity of the signal. This is shown to enhance the accuracy of sound eavesdropping by lowering the RMSE of frequency estimation from 44~Hz to 9~Hz. For the given data, we also note that further increase in the power of the interference beyond a certain point provides more quantization RSS values, but it will not improve accuracy below 9~Hz of RMSE.

\begin{figure*}[tbhp]

\begin{subfigure}[t]{0.48\textwidth}
     \includegraphics[width=0.76\columnwidth]{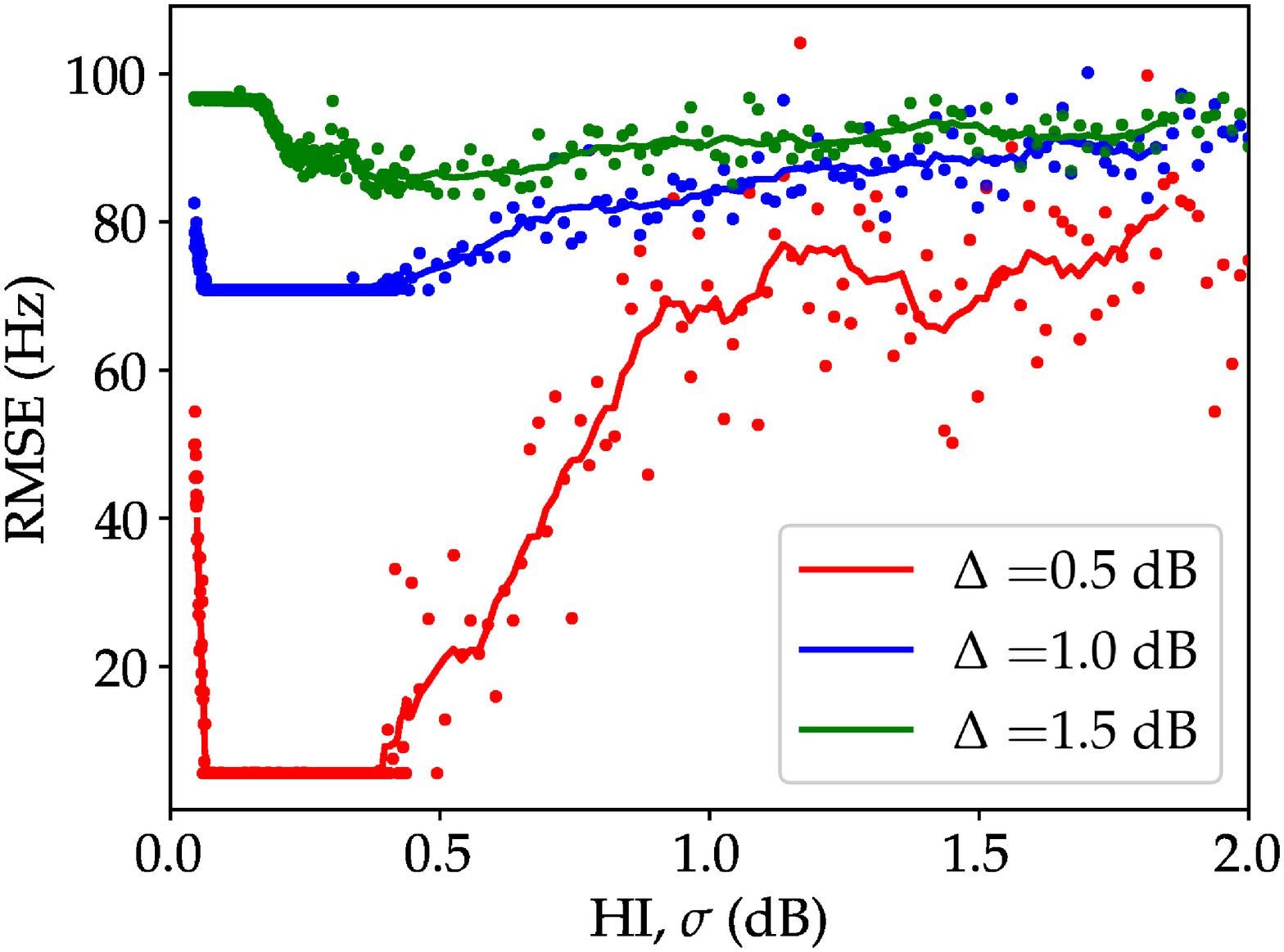}
     \caption{}
     \label{fig:rmse_vs_stepsize_noise}
 \end{subfigure} %
\quad
\begin{subfigure}[t]{0.48\textwidth}     
     \includegraphics[width=0.76\columnwidth]{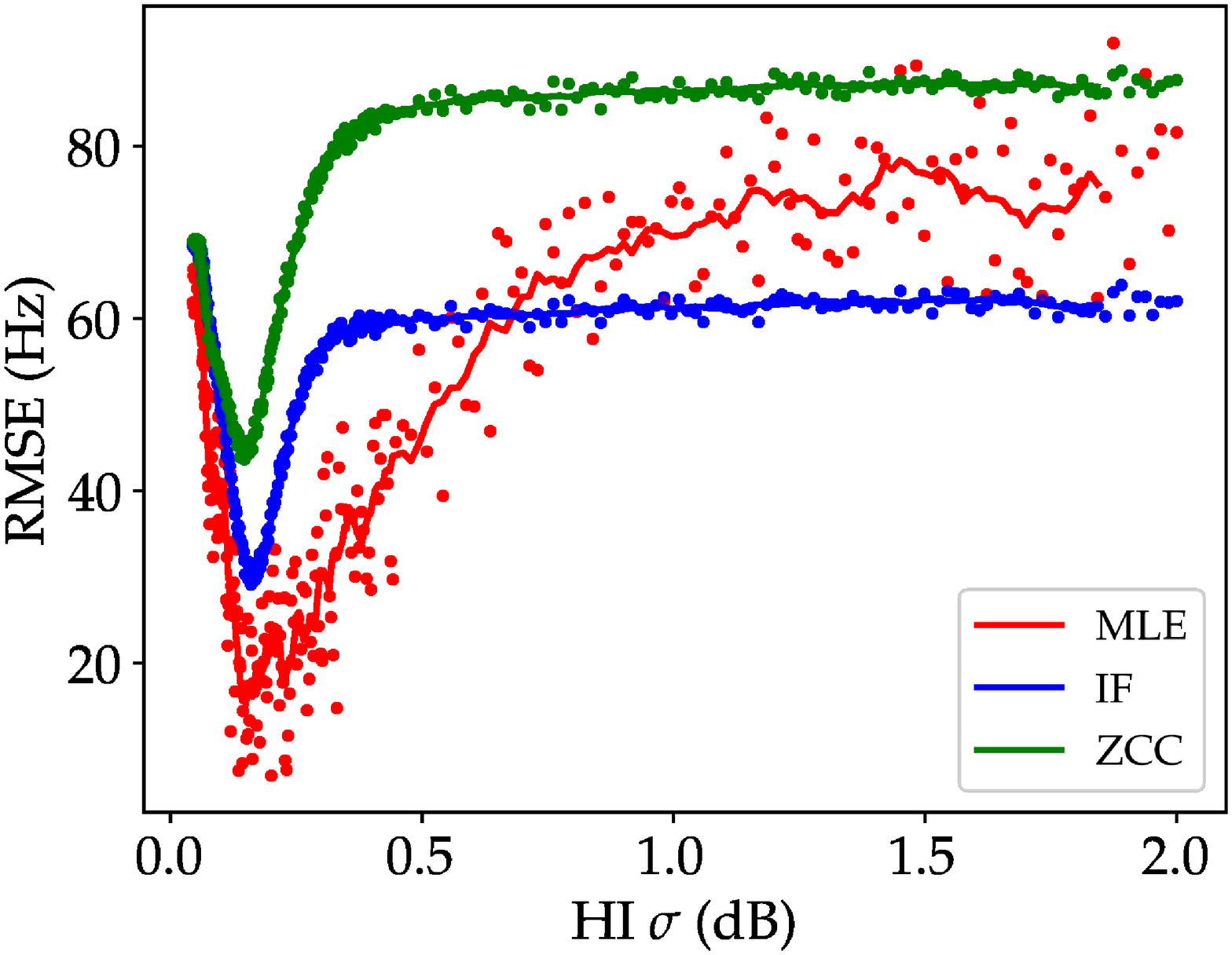}
    \caption{}
    \label{fig:rmse_vs_method_noise}
\end{subfigure} %
\caption{Sound frequency RMSE as a function of simulated interference when the RSS is quantized with different values of quantization step size $\Delta$ }
\label{fig:rmse_vs_noise}
\end{figure*}

{\bf Optimum HI.} From the result in Fig. \ref{fig:interference}, we note that there is a minimum value of the RMSE of sound frequency with respect to interference power for a given quantization step-size. In Fig. \ref{fig:rmse_vs_method_noise}, we observe that there exists an optimal interference power at which the RMSE of sound frequency gets its minimum for a given quantization step size. To study the relation between the optimum HI power and RSS quantization step-size, we compute the RMSE of sound frequency estimation for different quantization step-sizes.  In Fig. \ref{fig:rmse_vs_stepsize_noise}, the sound frequency RMSE of the MLE algorithm is given as a function of the standard deviation of HI when the received power is quantized with three different values of the quantization step size. 

\section{Bounds for RSS Surveillance}
\label{sec:bound}

The previous section provides experimental evidence of the possible benefits of HI, which an attacker can exploit to improve performance when RSS is quantized.
While the experimental results provide examples of what an attacker could achieve, they do not provide any guarantees about the best performance an attacker could achieve. 
In this section, we consider eavesdropping on single-tone sound, and provide analytical limits on the attacker's eavesdropping capability. These limits consider that an attacker may use HI, and are a function of the system parameters of available RSS sampling rate and quantization step size, as well as the amplitude of the sound signal being surveilled by the attacker. As before, we use the variance of frequency estimates of single-tone sound as an example.  Note however that the bound is generally applicable to any RF sensing application which estimates the amplitude or frequency of a sinusoidal signal component. We compute the theoretical lower bounds on estimation variance using Cram\'er-Rao bound analysis. 

The bounds on variance provide guarantees that are useful to both users and RFIC system designers.  First, note that one can never guarantee that an attacker cannot estimate the frequency of a sound tone at all --- an attacker can always estimate sound frequency to be 100 Hz, for example, without any RSS data, but this would not be a useful attack.  We focus on bounding the lowest possible variance of an attacker's unbiased sound frequency estimate since, if this variance is high, it effectively shows that the attacker is unable to gain meaningful information about the frequency of the sound.  A user could use such a bound to decide if an attacker who compromises the device could effectively monitor sound.  An RFIC designer could alter the parameters of the RSS measurements made available from the chip to increase the variance bound and thus make their device more acceptable to privacy-conscious customers.


\subsection{RSS Model for Single-tone Sound}
\label{ssec:model} 

In order to derive the theoretical bounds on RSS-based sound eavesdropping, we first model the received power including the variation due to single tone. As explained in \S \ref{sec:sound}, we assume that pure tone from speakers changes changes the measured signal as an additive sinusoidal component.  Here we again use \emph{received power} to indicate the continuous-valued power of the signal at the antenna, and RSS to indicate the quantized discrete-valued power reported by the transceiver IC.  Although generally an eavesdropper may take burst measurements, we assume a scalar time-dependant signal for simplicity.  We use $B$ to denote the received power when the sound is turned off, and $v(k)$ to denote the noise in sample $k$, which is assumed to be zero-mean white Gaussian noise with variance $\sigma^2$.  Therefore, the sampled received power signal is given as
\begin{equation} \label{eqn:model}
    x[k] = A\cos{(\omega T_s k+\phi)}+B + v[k],  \qquad k \in \mathbb{Z},
\end{equation}
where $T_s$ is the sampling period, and the sound-induced signal has unknown amplitude $A$, DC offset $B$, frequency $\omega$, and the initial phase $\phi$. The unknown parameter vector is  $\bm{\theta} = [A, B,\omega,\phi]^T$.  

Our model is that the received power is quantized with a step size of $\Delta$.  Typically $\Delta \gg A$, that is, the step size is significantly larger than the changes in RSS due to many RF sensing activities including breathing, pulse or sound. In our experimental study, involving a single tone audio played on Google Home Max speaker at maximum volume while a transceiver is placed on the same surface, we observe a peak-to-peak amplitude of 0.1~dB.  
Amplitudes of the breathing signal can be 0.1 dB \cite{luong2016rss}, or 0.3 dB \cite{uysal2017contactless}. Pulse-induced amplitudes are even smaller.  It is rare to see transceiver RSS to be quantized to less than 0.5 dB, as typical step sizes are 1.0 dB or higher.  Since $A$ is low compared to $\Delta$ the (quantized) RSS measurement typically takes one of two neighboring values.  It follows that we can approximate the RSS signal as a one-bit quantization of the received power $x[k]$. 
Note this approximation is not imperative for obtaining an estimation bound, since multi-bit CRB analysis of frequency analysis is possible \cite{moschitta2007cramer}.  When $\Delta \gg A$, that more complicated bound is nearly identical to the bound we derive, but the complexity can obscure the lessons learned from the bound with the one-bit quantization assumption.

Assuming one-bit quantization, the RSS is represented as:
\begin{equation} \label{eqn:quantization}
y[k] = {\rm sign}\left(x[k]-\zeta\right),
\end{equation}
where $\zeta$ is the threshold for quantization (the boundary between the two RSS bins) and the sign function ${\rm sign}(\cdot)$ is defined as ${\rm sign}(x)=1$ for $x\geq0$ and ${\rm sign}(x)=-1$ for $x < 0$. Without loss of generality, we assume $\zeta=0$.  The DC offset $B$ becomes the distance from the nearest quantization threshold and takes a value in the set $[-\Delta/2, \Delta/2]$. We define ${\bm y} = \left[y[0], \ldots, y[N-1]\right]^T$ to be our measurement vector.

The attacker's goal is to estimate $A$ and $\omega$ from these RSS measurements ${\bm y}$. oth sound amplitude and sound frequency provide vital information about the intensity and type of an activity in the surrounding. 




\subsection{Cram\'er-Rao Bound (CRB) Analysis}
\label{ssec:CRB}
Here, we compute the Cram\'er-Rao bound of the parameter vector ${\bm \theta}$ given the measurements ${\bm y}$. First we define the probability mass function corresponding to $k^{\mbox{th}}$ sample $y[k]$ as 
\[
f_{y[k]}=P(y[k]=q ; \, {\bm \theta}), \qquad q \in \{-1, +1\}.
\]
If we define ${\cal C}_k \coloneqq \cos (\omega T_s k+\phi)$ and ${\cal S}_k \coloneqq \sin (\omega T_s k+\phi)$, then
\begin{equation}
\begin{aligned}
f_{y[k]}(q;{\bm \theta}) &= P(y[k]=q;{\bm \theta}) =P(qx[k]>0;{\bm \theta})  \\ 
&= \frac{1}{\sqrt{2\pi}\sigma}\int^{\infty}_0 \exp\left({-\frac{[x-q(A{\cal C}_k+B)]^2}{2\sigma^2}}\right)dx.
\end{aligned} \nonumber
\end{equation}
Equivalently,
\begin{equation}
\begin{aligned}
f_{y[k]}(q;{\bm \theta})= \dfrac{1}{2}{\rm erfc}\left(-\dfrac{q}{\sqrt{2}\sigma}(A{\cal C}_k +B)\right).
\end{aligned}
\end{equation}
To compute the CRB, we first derive the Fisher information matrix (FIM).
From \cite{kay}, the element of the FIM from $i^{th}$ row and $j^{th}$ column is given by
\begin{equation}
\begin{aligned}
{\bm I}(\bm{\theta})_{ij} &= {\rm E}\left[  \dfrac{\partial \log f_{{\bm y}}(\bm{q};{\bm \theta})}{\partial \theta_i} \dfrac{\partial \log f_{{\bm y}}(\bm{q};{\bm \theta})}{\partial \theta_j}\right]\\
\end{aligned}
\end{equation}
Since the variables $y[k]$ are independent, the element of the FIM from $i^{th}$ row and $j^{th}$ column becomes:
\begin{equation} \label{eqn:fim_our_case}
\begin{aligned}
{\bm I}(\bm{\theta})_{ij} &= \sum_{k=0}^{N-1}\sum_{q=\pm 1} \frac{1}{f_{y[k]}(q;{\bm \theta})} \dfrac{\partial f_{y[k]}(q;{\bm \theta})}{\partial { \theta_i}} {\dfrac{\partial f_{y[k]}(q;{\bm \theta})}{\partial {\theta_j}}}
\end{aligned}
\end{equation}
We compute the partial derivatives for the parameters $\bm \theta$, plug them into (\ref{eqn:fim_our_case}), and the resulting FIM becomes:


\begin{equation}
\label{eqn:fim}
{\bm I}(\bm{\theta}) = \frac{2}{\pi \sigma^2} \sum_{k=0}^{N-1}
\dfrac{\exp{\left(-\frac{1}{\sigma^2}\left(A{\cal C}_k+B\right)^2\right)}}{1- {\rm erf}^2\left(\frac{1}{\sqrt{2}\sigma}(A{\cal C}_k +B)\right)}
{\bm F}_k,
\end{equation}
where

\begin{equation}
{\bm F }_k= 
\begin{bmatrix} 
{\cal C}_k^2  & {\cal C}_k & -AkT_s {\cal S}_k{\cal C}_k  & -A{\cal S }_k{\cal C}_k 
\\
{\cal C}_k    & 1        & -AkT_s {\cal S}_k   & -A{\cal S}_k 
\\
-AkT_s {\cal S}_k{\cal C}_k  & -AkT_s {\cal S}_k   & A^2 k^2T_s^2 {\cal S}_k^2  & A^2kT_s {\cal S}_k^2
\\
-A{\cal S}_k{\cal C}_k & -A{\cal S}_k & A^2  kT_s{\cal S}_k^2  &  A^2{\cal S}_k^2 
\end{bmatrix}. \nonumber
\end{equation}

In this analysis, we focus on finding the bounds on variance of unbiased amplitude ($\hat{A}$) and frequency ($\hat{\omega}$) estimators, which are given in the inverse of the FIM in (\ref{eqn:fim}) as
\begin{equation}
\begin{aligned}
\text {CRB}(\hat{A})&=\left\{{\bm I}(\bm{\theta})^{-1}\right\}_{11},  \\ 
\text {CRB}(\hat{\omega})&=\left\{{\bm I}(\bm{\theta})^{-1}\right\}_{33}.
\end{aligned}
\end{equation}

For a quantization step size $\Delta$, the DC offset $B$ can be represented as a uniform random variable defined over the interval $[-\Delta/2,\Delta/2]$. In addition, each bound is a weak function of the initial phase $\phi$ which is also random and uniform for our applications. Thus we average the CRB over uniform phase and uniform DC offset.  
We use $\overline{ \text {CRB}}$ to indicate the CRB averaged over a uniform phase $\phi$ and uniform DC offset $B$. Therefore, the variance of amplitude estimates $var(\hat{A})$ and the variance of frequency estimates $var(\hat{\omega})$ are bounded by $\overline{\text {CRB}}(\hat{A})$ and $\overline{\text {CRB}}(\hat{\omega})$ respectively.
\begin{equation}
\label{eqn:bound}
\begin{aligned}
var(\hat{A})&\geq \overline{\text {CRB}}(\hat{A})\\ 
var(\hat{\omega})&\geq \overline {\text {CRB}}(\hat{\omega}).
\end{aligned}
\end{equation}

In subsequent subsections, we study the accuracy of sound eavesdropping based on estimation variance computed in (\ref{eqn:bound}). In particular, we analyze the effects of quantization step size $\Delta$, interference $\sigma$ and sampling rate $f_s$ .

\subsection{Effects of Helpful Interference}
\label{ssec:noise}
We study the effect of adding noise to RSS measurements prior to quantization for both amplitude and frequency estimation. For this analysis, we consider low-frequency sounds particularly  $f$ = 100 Hz. Low frequency sounds like rumble noise in a car are common in our daily encounters.
  Further, we consider a sampling rate $f_s$ = 400~Hz. We set $N$, the number of samples, such that $NT_s = 1$ s, and we choose an amplitude $A=0.025$ dB.

\begin{figure*}[tbph]
  \begin{center}
  
     \includegraphics[width=0.4\textwidth]{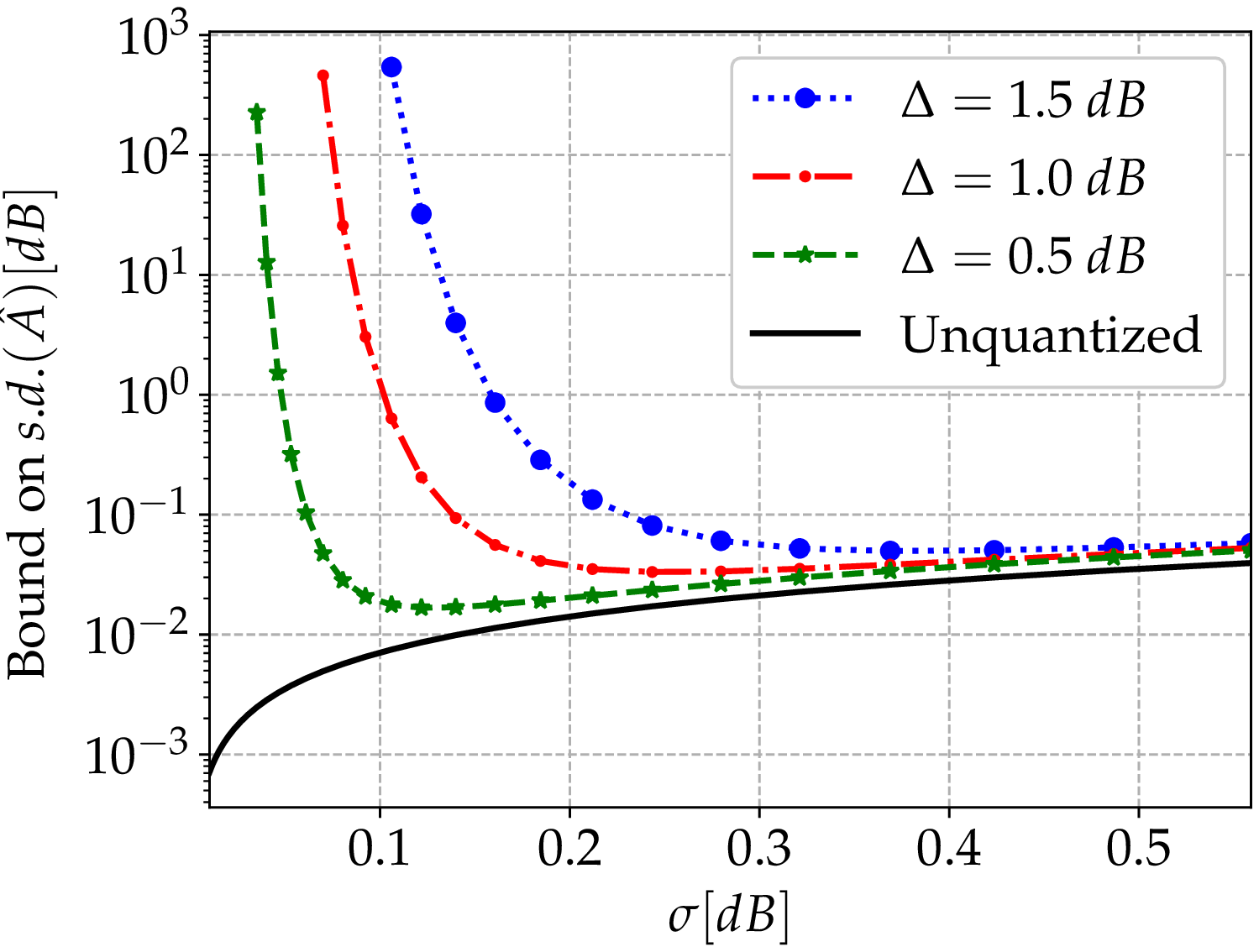}
  %
$\,\,\,\,\,\,\,\,$
  %
     \includegraphics[width=0.4\textwidth]{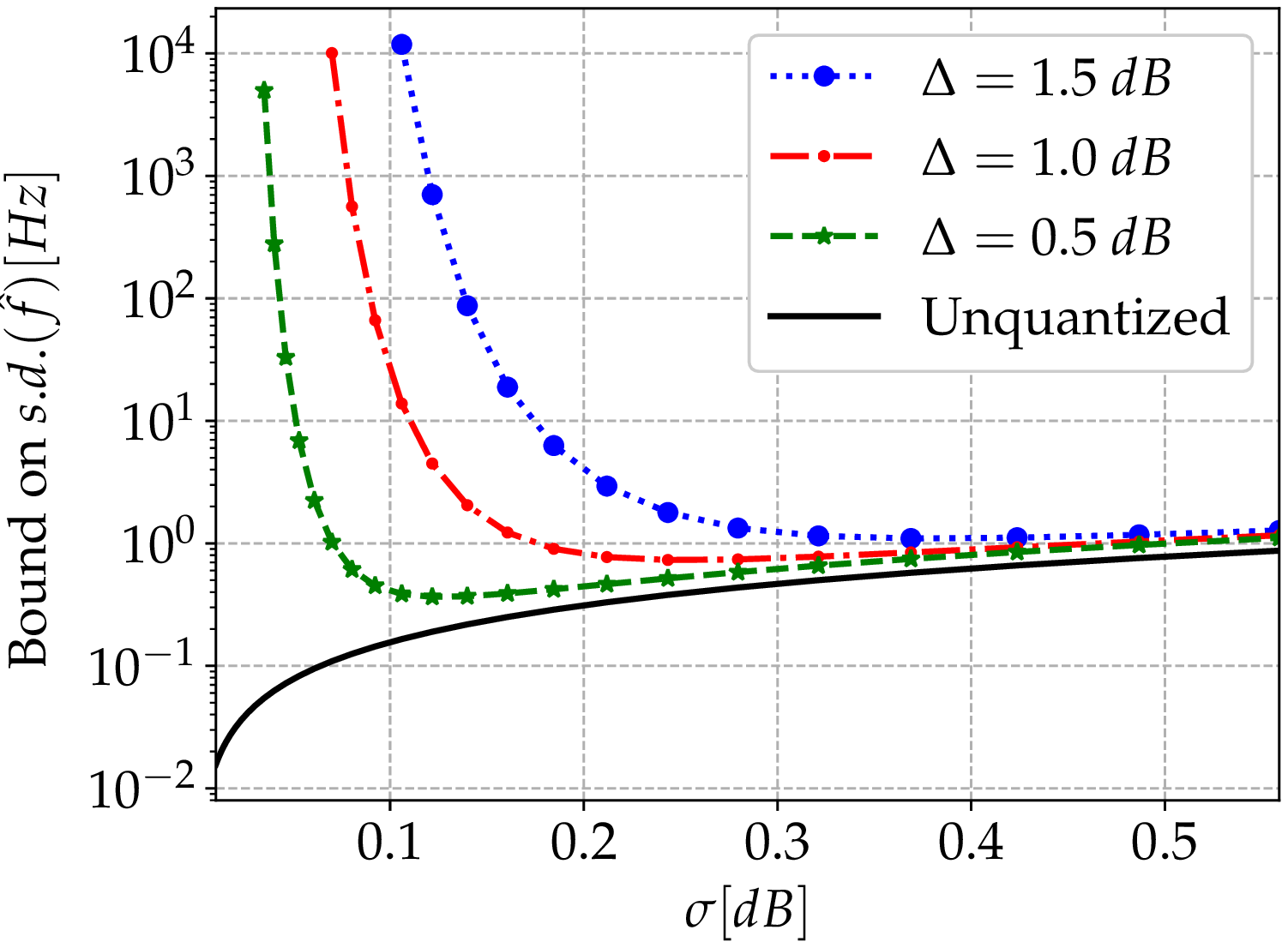}
%
\end{center}
  \caption{CRB of (Left) amplitude $A$ and (Right) frequency $\omega$, vs.\ noise power when $f_s = 400~Hz$, $f = 100~Hz$ and $A=0.025dB$. As noise power increases, the estimation variance decreases and then slowly increases.} \label{fig:noise_crb}
 \end{figure*}

We plot numerical results in Fig. \ref{fig:noise_crb} for the bound on the standard deviation of amplitude estimates as a function noise standard deviation as computed from (\ref{eqn:bound}).
We note that as the noise power increases, the bound on standard deviation of amplitude estimate generally decreases for quantized RSS measurements. Intuitively, this is because, as the sine wave is more likely further away from the threshold, even at its maxima or minima, estimation accuracy requires higher noise power in order to ensure that the measurements are not purely from one quantization level. For quantized RSS,  small interference power results in higher estimation variance as the quantized RSS measurements have a lower probability of changing their RSS levels with small noise power. 

This effect is similarly observed in frequency estimation. Fig. \ref{fig:noise_crb}(right) shows the effects of increasing noise power to RSS prior to quantization on the accuracy of sound frequency estimation. We see that increasing the noise level decreases the estimation variance for quantized RSS measurements. These results indicate that adding HI to a wireless channel improves the accuracy of amplitude and frequency estimations from quantized RSS measurements.  They also match the characteristics observed experimentally in \S \ref{sec:eval}.

It is also worthwhile to see the effect of adding noise when the signal is not quantized. We see the bound for $\hat{A}$ from \cite{rife1974single}, is $\mbox{var}(\hat{A}) \ge 2\sigma^2/N$, which indicates that the standard deviation increases with noise power for unquantized RSS measurements. Similarily,  we see that increasing the noise level increases the estimation variance of frequency estimates for unquantized RSS measurements. 

\vspace{0.1in}

\begin{figure}[h]
  \begin{center}
  \includegraphics[width=0.9\columnwidth]{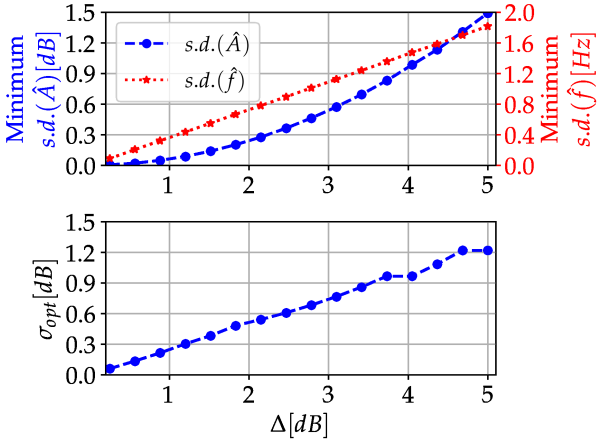}
   \end{center}
  \caption{CRB as a function of quantization step size}\label{fig:stepsize_bound}
 \end{figure}

\noindent  {\bf Optimum Noise Variance}:
A key observation from the results is that the bound on estimation variance using quantized measurements has a minimum value with respect to noise power for a given sampling rate and quantization step size. 
We note from Fig. \ref{fig:noise_crb} that there exists an optimal noise level at which estimation variance is brought to its minimum for a given sampling rate and RSS quantization step size. 
Our numerical results show that the optimal noise level for amplitude estimation matches that for frequency estimation. In Fig. \ref{fig:stepsize_bound}, we observe that the standard deviation of this optimal noise is linearly proportional to the RSS quantization step size, and that $\sigma_{opt}$ is approximately $\Delta / 4$.  
Interestingly, this standard deviation of helpful interference is just less than the standard deviation of quantization error, which is $\Delta/\sqrt{12} = \Delta / 3.46$.

\begin{figure}[btph]
\begin{minipage}[h]{0.95\linewidth}
\centerline{\includegraphics[width=0.95\columnwidth]{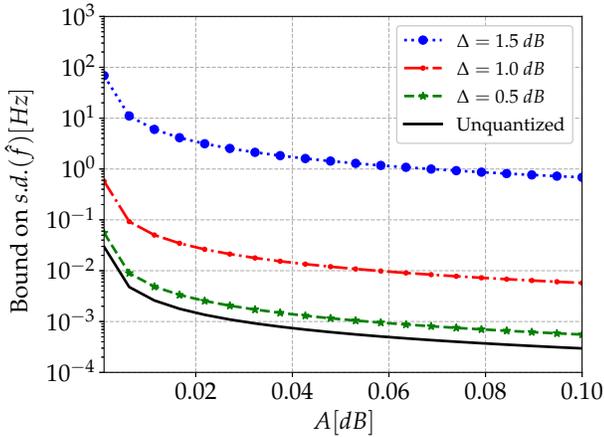}}
\end{minipage}
\caption{Effect of amplitude on frequency estimation}
\label{fig:ampf}
\end{figure}

It should be noted that the bound on standard deviation of $\hat{\omega}$ is a weak function of the frequency parameter $\omega$, and thus the plot is omitted. However, the amplitude significantly affects the performance of frequency estimation; as shown in Fig. \ref{fig:ampf}, higher amplitude results in lower standard deviation of frequency estimates.
 
 \subsection{Effects of Quantization Step Size}
An other parameter that controls the performance of sound eavesdropping is the quantization step size $\Delta$. 
We can see that the accuracy of sound eavesdropping, despite the ability of the attacker to use helpful interference, can be generally be degraded by increasing the RSS quantization step size. Furthermore, the minimum estimation bounds for amplitude and frequency estimates behave differently with respect to the RSS quantization step size. In Fig. \ref{fig:stepsize_bound}(top), we observe that the bound for frequency estimates increases linearly with RSS quantization step size whereas the bound for amplitude estimate fits a quadratic model with respect to the step size.

\begin{figure*}[phtb]
\begin{center}
     \includegraphics[width=0.4\textwidth]{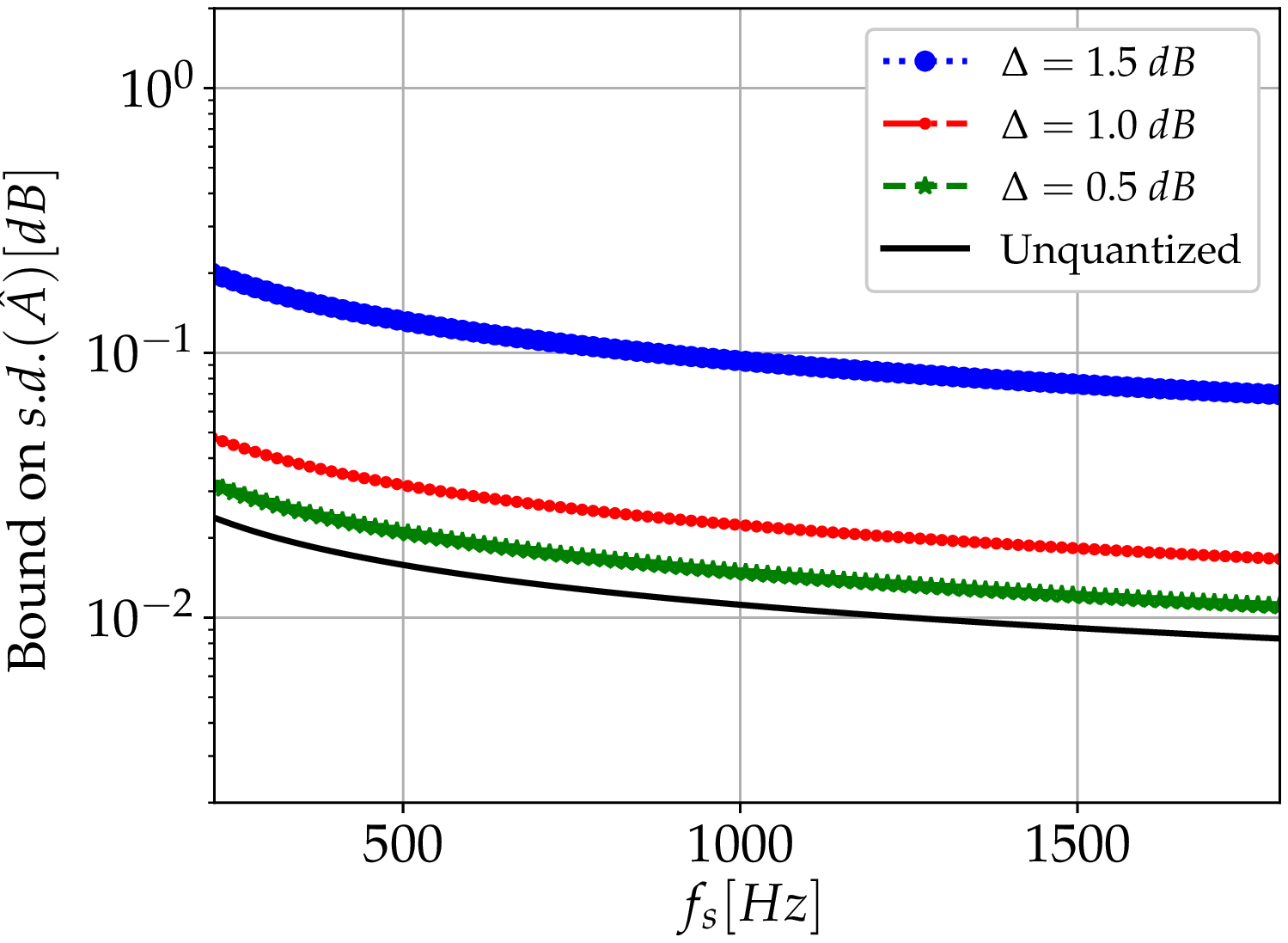}
  %
$\,\,\,\,\,\,\,\,$
  %
     \includegraphics[width=0.4\textwidth]{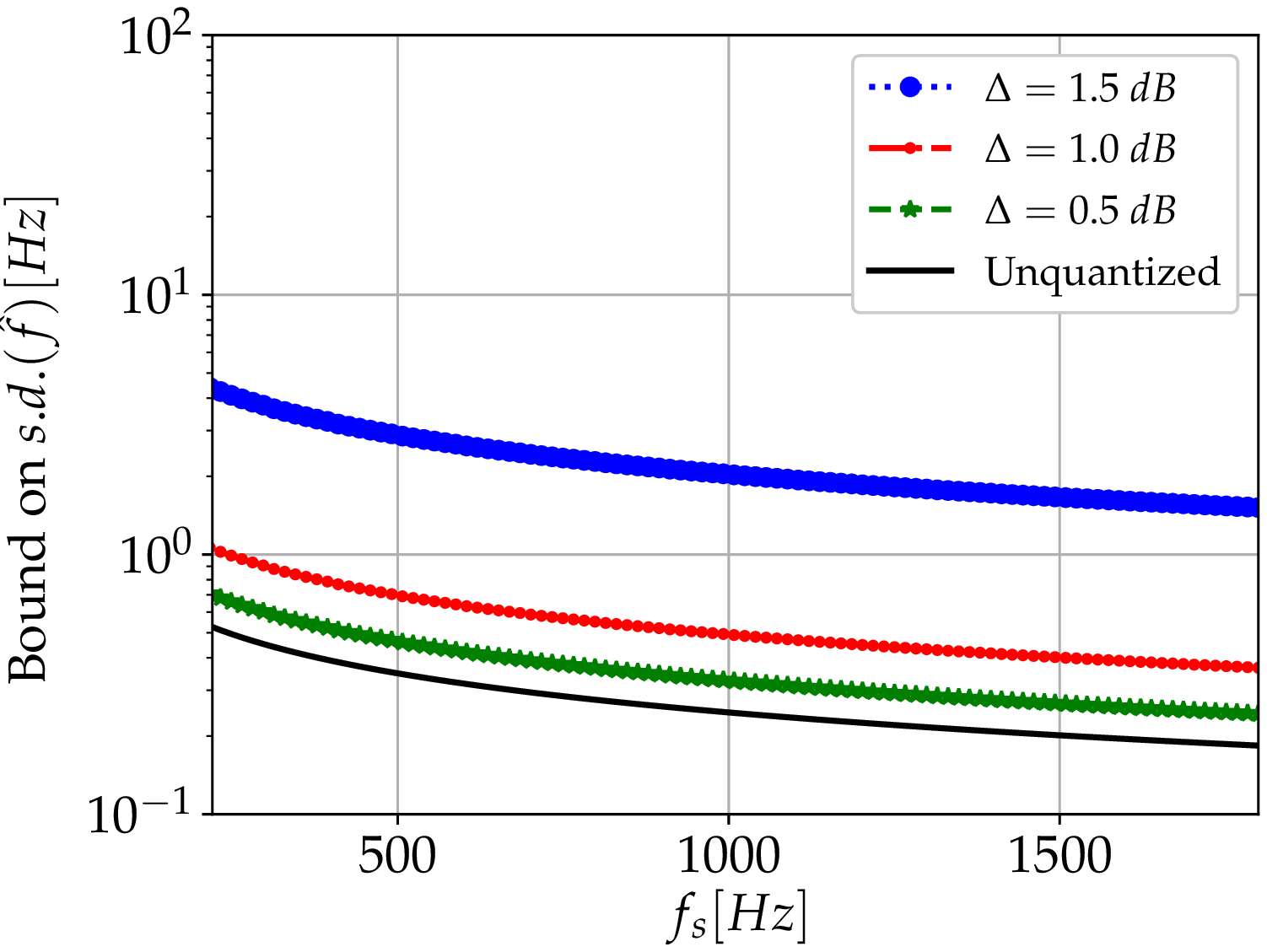}
%
\end{center}
  \caption{CRB of (Left) amplitude $A$ and (Right) frequency $\omega$, as a function of sampling rate when $\sigma = 0.25~dB$. As sampling rate $f_s$ increases, the estimation variance decreases.}
 \label{fig:sample_crb}
 \end{figure*}

\subsection{Effects of RSS Oversampling}
\label{ssec:oversample}
Next, we evaluate the effects of sampling rate on the accuracy sound eavesdropping, particularly in frequency estimation. We use $\omega=100$ Hz, and $A=0.025$ dB. 
In Fig.~\ref{fig:sample_crb}(left), we plot the bound on the standard deviation of amplitude estimate as a function of the sampling rate $f_s$. This bound decreases monotonically with $f_s$ for any value of the quantization step size $\Delta$.  The lowest $f_s$ in Fig.~\ref{fig:sample_crb}(left) is 1 Hz.  This suggests that an eavesdropper attains lower estimation variance by collecting RSS at higher rate. If it is possible to increase the sampling rate, the bound shows the possibility of order-of-magnitude decreases in standard deviation. Similar results are observed for frequency estimation where increasing sampling rate decreases the bound on standard deviation of frequency estimates.

\subsection{Overall Effects}
\begin{figure*}[tbph]
  \begin{center}
  
     \includegraphics[width=0.4\textwidth]{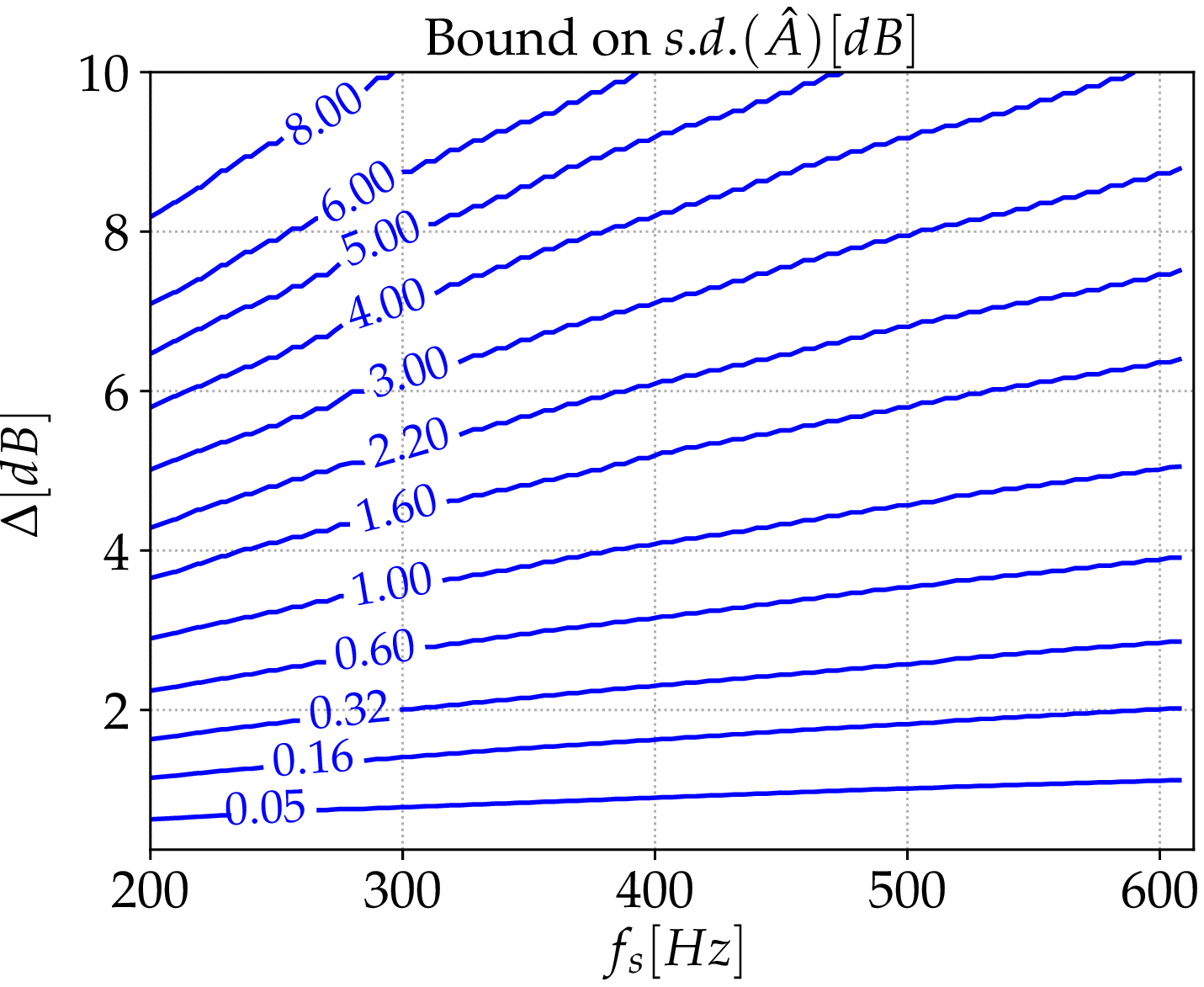}
  %
$\,\,\,\,\,\,\,\,$
  %
     \includegraphics[width=0.4\textwidth]{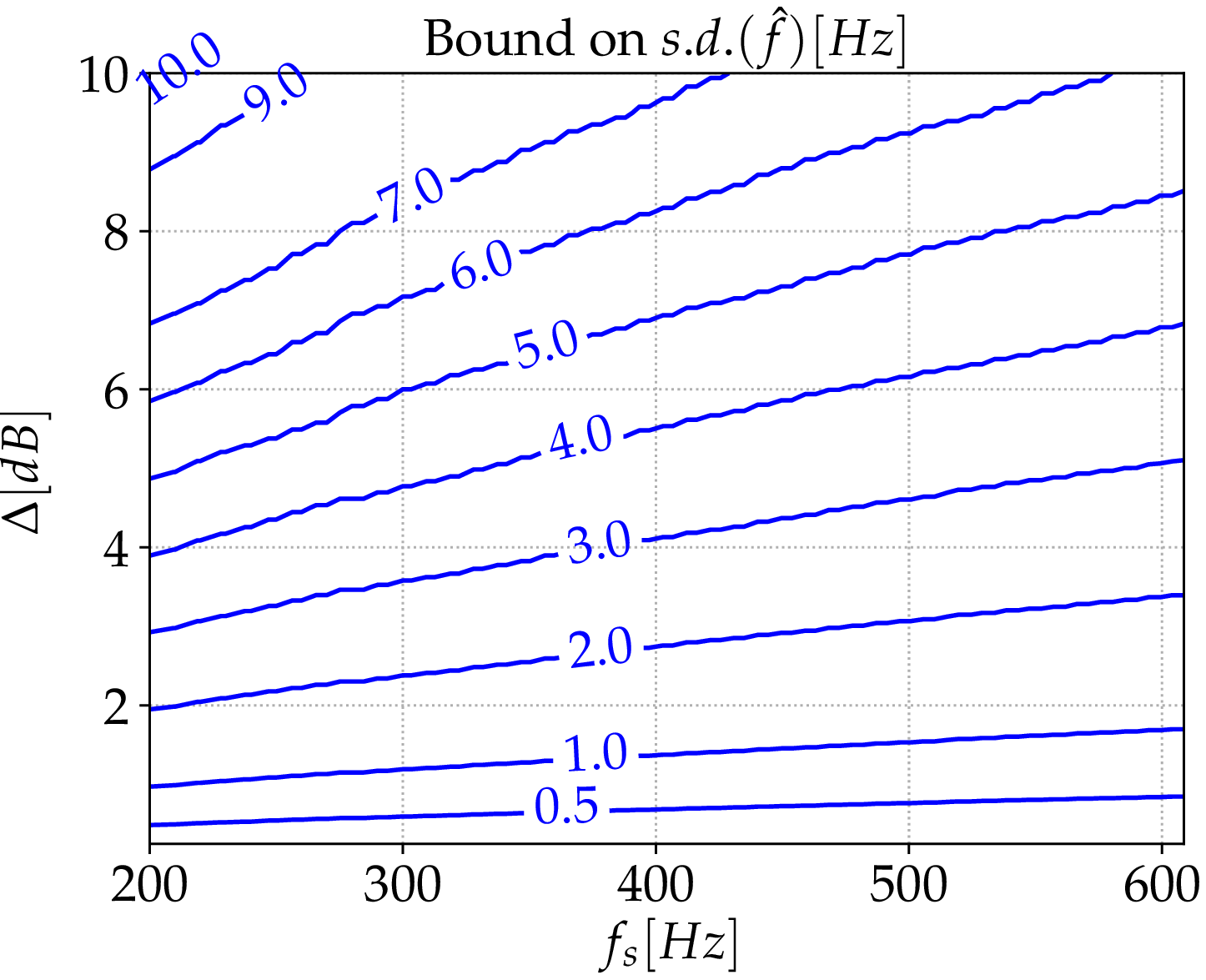}
\end{center}
  \caption{Contour plot of minimum CRB for (Left) amplitude and (Right) frequency estimates vs.\ sampling rate and quantization step size for a 100~Hz signal with amplitude of 0.025~dB.} 
  \label{fig:sample_noise}
 \end{figure*}

Our CRB analysis shows that the accuracy of RSS-based surveillance can be controlled mainly by two parameters: RSS quantization step size and sampling rate at which the RSS measurement is collected.  
We numerically analyze the combined effect of oversampling and quantization on RSS-based sound eavesdropping. Fig. \ref{fig:sample_noise}(left) shows the lower bound standard deviations of amplitude and frequency estimates as functions of quantization step size $\Delta$ and RSS sampling rate $f_s$. We note from these plots that lower estimation variance is generally attained with higher sampling rate and low step size. On the other hand, lower sampling rate and high step size lead to large estimation variance and hence lower accuracy in sound surveillance. For example, For a 100 Hz signal with A=0.025~dB, a quantization step size of 4~dB and sampling rate of 400~Hz provides 1~dB as the minimum standard deviation in amplitude estimates, which is too large compared to the given amplitude. 
The paper presents the sound eavesdropping capability of RSS measurements more than previously known. Prior research in sound eavesdropping relies on fine-grained measurements from software-defined radio platforms with measurements \cite{wei2015acoustic}. However, standard RSSI measurements are quantized with large step sizes which deters sound eavesdropping when there is no HI. 

We show both theoretically and experimentally that reliable sound eavesdropping could be obtained using helpful interference.  An attacker with knowledge of the quantization step size $\Delta$ could force one or more devices to transmit HI to obtain the optimum interference at the highest sampling rate possible and achieve reliable sound eavesdropping.
These results suggest the accuracy of an RSS-based sound surveillance attack can be limited by selecting a large RSS step size and low sampling rate. A designer could take sampling rate and quantization into account for systems with critical privacy requirements.

The analysis presented in this article considers single-tone sound vibrations; however the same mathematical framework can be extended to study RSS-based eavesdropping of sound with multiple tones, and other periodic signals such as pulse and respiration \cite{abrar2019quantifying}.

\section{Related work}
\label{sec:related}

{\vspace{0.1in} \noindent \bf Radio Frequency Sensing:}  
Radio Frequency sensing uses radio channel measurements to monitor human vital signs, detect activity or monitor sound in the environment. 
Various radio channel measurements such as received signal strength (RSS), channel impulse response (CIR), and channel state information (CSI) have been used for multiple RF sensing applications including contact-free vital sign monitoring \cite{patwari2014breathfinding,liu2015tracking}, device-free localization \cite{kaltiokallio2012follow, yang2013rssi}, gesture and activity recognition \cite{wang2015understanding, sigg2014telepathic}, and human identification \cite{xin2016freesense}. 

Among several channel measurements employed in most commercial wireless systems, RSS is considered to be the most widely available measurement across diverse wireless platforms \cite{liu2017wireless}. 
RSS has been applied in various RF sensing applications including acoustic eavesdropping \cite{wei2015acoustic}, device-free localization \cite{youssef2007challenges}, contact-free vital sign monitoring \cite{abrar2019pulse, patwari2014breathfinding}, security \& surveillance \cite{hussain2008using}, activity and gesture recognition \cite{sigg2014rf,luong2016rss}, and home monitoring \cite{kaltiokallio2012follow}. 
In \cite{wei2015acoustic}, signal strength measurements using a software-defined platform are used to eavesdrop acoustic vibrations from large speakers. 


The ease of access to RSS and its capability in RF sensing allows a potential threat on privacy. Little attention has been paid to these threats, mainly because RSS-based sensing has been reported to have limited reliability as a result of its relatively large quantization step size \cite{luong2016rss}. However, the limits on the capability of RSS-based sensing has not been fully explored. In this paper, we experimentally demonstrate such an unexplored RF sensing capability that uses noise superimposed in RSS measurements to improve sound eavesdropping.  

{\vspace{0.1in} \noindent \bf Estimation Bounds:}  
Estimation bounds are statistical tools used to evaluate the performance of algorithms in estimating certain parameters of interest with respect to the maximum theoretically attainable accuracy, commonly based on their estimation variance. 
The Cram\'er Rao lower bound is the most common variance bound due to its simplicity \cite{kay}.  It provides the lowest possible estimation variance achieved by any unbiased estimator.

Evaluating the accuracy of algorithms used in RSS surveillance is essential step to determine eavesdropper's capabilities. 
An analytical explanation for the relationship between noise power, sampling rate, amplitude, quantization, and parameter estimator performance has not been presented. 
The change in RSS due to periodic activities like respiration can be generally modelled as a sine wave \cite{patwari2014monitoring}.  
For unquantized sine wave signal, the CRB on the variance of unbiased frequency estimators is derived in \cite{kay, rife1974single}. 
However, RSS has a signficantly higher quantization step size than the typical RSS change induced by vital signs, each RSS sample primarily falls into either of two successive RSS values. 
H{\o}st-Madsen \emph{et al.} \cite{host2000effects} quantitatively explain the effect of quantization and sampling on frequency estimation of a one-bit quantized complex sinusoid, 
but without presenting bounds for amplitude estimation or considering a DC offset as a complicating parameter.  In this paper, we evaluate attacker's bound on breathing rate and amplitude estimation in the realistic case that there is a DC offset.  Further we have demonstrated what is observed in the bound, that increased interference power can actually help improve estimates.

{\vspace{0.1in} \noindent \bf Wireless Network Security:}  
Security in wireless networks is conventionally achieved through cryptographic protocols at multiple layers in the network stack. In wireless local area networks including 802.11, a number of cryptographic protocols have been standardized including IPSecurity, 
 Wi-Fi Protected Access (WPA), and Secure Sockets Layer (SSL). 
Due to the broadcast nature of the wireless medium, researchers have proposed additional security protocols at the physical layer to deter eavesdropping and jamming, such as by exploiting channel characteristics \cite{tomko2006physical, li2005mimo}, employing coding schemes \cite{shiu2011physical}, or controlling signal power \cite{noubir2004connectivity,gollakota2011they}. 
However, these approaches do not prevent an adversary already with some access to a system from using PHY layer signal measurements for sensing purposes. Moreover, even if the end-to-end encryption prevents the attacker access to the data from the source the attacker can still access the RSS from received packet.


Such an adversary can also force a wireless device to continuously transmit helpful interference in order to reduce the effect of quantization on the quality of the RSS information. Most wireless standards use a multiple access control method to avoid interference, such as carrier-sense multiple access (CSMA). However, many RFICs (e.g., Atheros AR9271) provide the option to disable CSMA and control the random backoff timer \cite{vanhoef2014advanced}. These vulnerabilities pave the way for the attacker to change a device's software to create an interferer operating on the same channel at the same time as the receiver measuring RSS.

Despite considerable research in privacy, an RSS surveillance attack exploiting measurements from wireless systems is an unresolved problem. 
Banerjee \emph{et al.} demonstrate  that an attacker can easily estimate artificial changes in transmit power to detect and locate people through a wall. In \cite{qiao2016phycloak}, an third device is introduced to distort the PHY layer signal before it could be used by eavesdropper for sensing purposes, but fails for multiple-antenna eavesdropper or if a device can be remotely compromised and caused to run the attacker's software.

\section{Conclusion}
\label{sec:conc}

In this paper, we explore the limits on RSS-based eavesdropping of sound vibrations.  We analyze the capability of an attacker in estimating the sinusoidal parameters of low-amplitude sinusoidal signals by deriving the theoretical lower bound with which an attacker could estimate the rate and amplitude of a sinusoid.  We show, both theoretically and experimentally, that the adversary could force other wireless devices to transmit simultaneously in order to improve their estimates.  
The numerical values of the lower bound on variance show, for typical RFICs, an RSS-surveillance attack could be very accurate.
We discuss, as a result, how a device designer could limit the performance of a potential attack by adjusting the quantization step size and the sampling rate. Most commercial transceivers have fixed RSS quantization schemes, however, a manufacturer could adjust RSS quantization to ensure that sound eavesdropping attacks are ineffective.

\appendices
\section{Proof for Partial Derivatives}
\label{app:partial}

Given ${\cal C}_k \coloneqq \cos (\omega T_s k+\phi)$, ${\cal S}_k \coloneqq \sin (\omega T_s k+\phi)$ and $q\in\{-1, +1\}$, then

\begin{equation}
\begin{aligned}
\dfrac{\partial}{\partial A} f_{y[k]}(q;{\bm \theta})&= \frac{1}{2}\dfrac{\partial}{\partial A} {\rm erfc}\left(-\dfrac{q}{\sqrt{2}\sigma}(A{\cal C}_k +B)\right)\\
&= \frac{1}{\sqrt{\pi}} \exp{\left(-\frac{1}{2\sigma^2}\left(A{\cal C}_k+B\right)^2\right)}\\
&\qquad\quad \dfrac{\partial}{\partial A}\left(-\dfrac{q}{\sqrt{2}\sigma}(A{\cal C}_k +B)\right)\\
&= -\frac{q{\cal C}_k}{\sqrt{2\pi} \sigma} \exp{\left(-\frac{1}{2\sigma^2}\left(A{\cal C}_k+B\right)^2\right)}
\end{aligned}
\end{equation}

\begin{equation}
\begin{aligned}
\dfrac{\partial}{\partial B} f_{y[k]}(q;{\bm \theta})&= \frac{1}{2}\dfrac{\partial}{\partial B} {\rm erfc}\left(-\dfrac{q}{\sqrt{2}\sigma}(A{\cal C}_k +B)\right)\\
&= \frac{1}{\sqrt{\pi}} \exp{\left(-\frac{1}{2\sigma^2}\left(A{\cal C}_k+B\right)^2\right)}\\
&\qquad\quad \dfrac{\partial}{\partial B}\left(-\dfrac{q}{\sqrt{2}\sigma}(A{\cal C}_k +B)\right)\\
&= -\frac{q}{\sqrt{2 \pi} \sigma} \exp{\left(-\frac{1}{2\sigma^2}\left(A{\cal C}_k+B\right)^2\right)}
\end{aligned}
\end{equation}

\begin{equation}
\begin{aligned}
\dfrac{\partial}{\partial \omega} f_{y[k]}(q;{\bm \theta})&= \frac{1}{2}\dfrac{\partial}{\partial \omega} {\rm erfc}\left(-\dfrac{q}{\sqrt{2}\sigma}(A{\cal C}_k +B)\right)\\
&= \frac{1}{\sqrt{\pi}} \exp{\left(-\frac{1}{2\sigma^2}\left(A{\cal C}_k+B\right)^2\right)}\\
&\qquad\quad \dfrac{\partial}{\partial \omega}\left(-\dfrac{q}{\sqrt{2\sigma}}(A{\cal C}_k +B)\right)\\
&= \frac{qAT_sk{\cal S}_k}{\sqrt{2\pi} \sigma} \exp{\left(-\frac{1}{2\sigma^2}\left(A{\cal C}_k+B\right)^2\right)}
\end{aligned}
\end{equation}

\begin{equation}
\begin{aligned}
\dfrac{\partial}{\partial \phi} f_{y[k]}(q;{\bm \theta})&= \frac{1}{2}\dfrac{\partial}{\partial \phi} {\rm erfc}\left(-\dfrac{q}{\sqrt{2}\sigma}(A{\cal C}_k +B)\right)\\
&= \frac{1}{\sqrt{\pi}} \exp{\left(-\frac{1}{2\sigma^2}\left(A{\cal C}_k+B\right)^2\right)}\\
&\qquad\quad \dfrac{\partial}{\partial \phi}\left(-\dfrac{q}{\sqrt{2}\sigma}(A{\cal C}_k +B)\right)\\
&= \frac{qA{\cal S}_k}{\sqrt{2\pi} \sigma} \exp{\left(-\frac{1}{2\sigma^2}\left(A{\cal C}_k+B\right)^2\right)}
\end{aligned}
\end{equation}

\section*{Acknowledgement}
This material is based upon work supported by the US Army Research Office under Grant No. 69215CS.

\balance
\bibliographystyle{IEEEtran}
\bibliography{main.bbl}


\end{document}